\documentclass[preprint,showpacs,preprintnumbers,amsmath,amssymb]{revtex4}
  
\usepackage[dvipdfmx]{graphicx}% Include figure files
\usepackage{dcolumn}% Align table columns on decimal point
\usepackage{bm}% bold math
 
\newcommand{\be}{\begin{equation}}
\newcommand{\ee}{\end{equation}}
\newcommand{\eq}[1]{Eq.~(\ref{#1})}
\newcommand{\fig}[1]{Fig.~\ref{#1}}
\def\bea{\begin{eqnarray}}
\def\eea{\end{eqnarray}}

\def\bra{\langle}
\def\ket{\rangle}
\def\vq{{\bf q}}

\def\vk{{\bf k}}
\def\vQ{{\bf Q}}

\begin{document}
 
\title{ 
Static spin susceptibility in magnetically ordered states
}

% Force line breaks with \\ 

\author{Kazuhiro Kuboki}
%\email{kuboki@kobe-u.ac.jp}
\affiliation{%
Department of Physics, Kobe University, Kobe 657-8501, Japan
}%
\author{Hiroyuki Yamase}
%\email{yamase.hiroyuki@nims.go.jp}
\affiliation{National Institute for Materials Science, Tsukuba 305-0047, Japan 
}%

\date{\today}

\begin{abstract}
We report that special care is needed when longitudinal magnetic susceptibility 
is computed in a magnetically ordered phase, especially in metals. We demonstrate  
this by studying static susceptibility in both a ferromagnetic and an 
antiferromagnetic state 
in the random phase approximation to the two-dimensional Hubbard model on a square lattice. 
In contrast to the case in the disordered phase, a first derivative of the chemical potential (or the density) 
with respect to a magnetic field does not vanish in a magnetically ordered phase 
when the field is applied parallel to the magnetic moment. 
This effect is crucial and should be included when computing  magnetic susceptibility in the ordered phase, 
otherwise an unphysical result would be obtained.  
In addition, consequently the magnetic susceptibility becomes different when computed at a 
fixed density and a fixed chemical potential in the ordered phase.  
In particular, we cannot employ magnetic susceptibility at a fixed chemical potential 
to describe a system with a fixed density even if the chemical potential is tuned 
to reproduce the correct density. 
%%%% added %%%%
%For an insulating magnetic state, contributions from first 
%derivatives of the chemical potential or density should vanish 
%due to a charge gap. Hence special care may not be needed
%in contrast to the case of metallic magnetic states.   
%%%%%%%%%%%
\end{abstract} 

\pacs{
}

\maketitle
\section{Introduction} 
Spin susceptibility is a fundamental quantity to study the magnetic property of 
a system, and it 
is often computed in the so-called random phase approximation (RPA). 
While this approximation is usually good enough for a three-dimensional system, 
it may not be precise enough especially in a two-dimensional system. However, even in such a case, 
the susceptibility computed in the RPA is believed to capture at least qualitative properties of the system. 
 
RPA susceptibility is frequently computed in a disordered phase, but it 
can also be computed  in a magnetically ordered phase \cite{izuyama63,ueda78}. 
Moreover, as actually observed in high-temperature cuprates \cite{mukuda12}, 
iron-based pnictides and chalcogenides \cite{stewart11}, 
and heavy fermion materials \cite{pfleiderer09}, 
the ordered phase sometimes coexists with superconductivity. 
Even in such a complicated situation, the RPA provides feasible computations of 
magnetic susceptibility \cite{hjlee12,rowe12}. 

Typically, RPA susceptibility is obtained by connecting a simple bubble (or ladder) of noninteracting 
particle-hole excitations with the electron-electron interaction, that is, 
its functional form is given typically by 
\be
\chi \propto  (1 - \chi_0 g)^{-1} \chi_0  \,,
\label{usualRPA}
\ee
where $g$ is the interaction strength and $\chi_0$ is the susceptibility in the noninteracting case; 
$\chi$, $\chi_0$, and $g$ can be matrices. 
In a magnetic phase, $\chi_0$ is computed by using 
the quasiparticle propagator in the ordered phase, and also by considering possible  umklapp 
contributions to the susceptibility when the translational symmetry is broken by a magnetic order. 
Such a procedure indeed yields the correct result of transverse magnetic susceptibility 
\cite{izuyama63,ueda78,schrieffer89,chubukov92,kampf94,knolle10,hjlee12,rowe12}, but 
special care is needed for longitudinal magnetic susceptibility, which is not 
well recognized \cite{sokoloff69,schrieffer89, kampf94, knolle10,rowe12}. 

Spin rotational symmetry is broken in a magnetically ordered phase. 
As a result, the chemical potential (or the density) is no longer a quadratic function 
of a magnetic field when the field is applied parallel to the magnetic moment. 
A first derivative of the chemical potential (or the density) then becomes finite. 
Hence this effect should be considered on an equal footing when 
we compute  magnetic susceptibility, because the magnetic susceptibility 
is a linear-response quantity of a magnetic field. 

In this paper, we show how important the contribution of the first derivative 
of the chemical potential (or the density) is to compute longitudinal susceptibility 
in a magnetically ordered phase, 
which we exemplify by employing the two-dimensional Hubbard model for both 
a ferromagnetic 
and an antiferromagnetic state. Since the RPA is equivalent to the 
mean-field approximation or the saddle-point approximation, 
we can directly compute the magnetic susceptibility in mean-field theory for the Hubbard model. 
We provide the correct expression of the static susceptibility in the RPA 
as well as results when the first derivative of the chemical potential (or the density) is 
neglected. In addition, we point out that the longitudinal magnetic susceptibility is different 
when computed at a fixed density and a fixed chemical potential 
in a magnetically ordered phase. 
Consequently, when the density is fixed, the susceptibility obtained at a fixed chemical potential 
cannot be applicable even if the chemical potential is tuned 
to reproduce the correct density. 

This paper is organized as follows. 
In Sec. II we present the model and derive the self-consistency equations 
for both a ferromagnetic and an antiferromagnetic phase. 
The corresponding magnetic susceptibility is computed in Secs. III and IV, respectively.  
We show in Sec. V that the susceptibility for a fixed chemical potential is reproduced 
in a conventional diagrammatic approach. Concluding remarks are given in Sec. VI.

\section{Model and self-consistency equations}
To exemplify our issue, we employ the two-dimensional Hubbard model on a square lattice, 
\be
\mathcal{H}= -\sum_{i,j,\sigma} t_{ij} c^\dagger_{i \sigma} c_{j \sigma}
 +U\sum_j n_{j \uparrow} n_{j \downarrow} + \mathcal{H}_z, 
\label{Hubbard}
 \ee
where the transfer integrals $t_{ij}$ are finite between the first- ($t$) 
and second- ($t'$) neighbor sites and otherwise zero; 
$U$ represents the on-site Coulomb repulsion. 
$\mathcal{H}_z$ is the Zeeman term, for which 
we consider a static and uniform (staggered) magnetic field when we compute 
longitudinal magnetic susceptibility in a ferromagnetic (an antiferromagnetic) state. 
That is, it is described as 
\be
\mathcal{H}_z = -h \sum_j \frac{1}{2} (n_{j \uparrow}-n_{j \downarrow})
e^{{\rm i}{\bf q}\cdot {\bf r}_j}
\label{zeeman}
\ee
with $\vq=0$ [$\vq={\vQ} \equiv (\pi, \pi)$]. 
Here $h$ is an effective magnetic field given by $h=\mathfrak{g}\mu_{B}H$; 
$\mathfrak{g}$ is a $g$ factor, $\mu_B$ the Bohr magneton, and $H$ an external magnetic field. 
The magnetic field is infinitesimally small and we take the limit of $h\rightarrow +0$ 
when we compute the susceptibility.  

Since the RPA is equivalent to the mean-field approximation, we compute 
the RPA susceptibility in mean-field theory. 
Defining the magnetization and the density 
operator as 
\bea
&&m_j=\frac{1}{2} (n_{j \uparrow} - n_{j \downarrow}) \,,\\
&&n_j=n_{j \uparrow} + n_{j \downarrow} \,,
\eea
respectively, the interaction term is written as $n_{j\uparrow} n_{j\downarrow} =\frac{1}{4} n_j n_j - m_j m_j$. 
The density is assumed to be uniform and is given by $n=\bra n_j \ket$ whereas 
the magnetization $\bra m_j \ket$ 
is uniform in the ferromagnetic state 
and staggers with a wavevector $\vq=\vQ$ in the antiferromagnetic state. 
In mean-field theory the interaction term is decoupled as 
\be
n_{j\uparrow} n_{j\downarrow} \rightarrow \frac{n}{2} n_j - 2 \bra m_j \ket m_j - \frac{1}{4} n^2 + \bra m_j \ket^2 \,,
\ee
and self-consistency equations for $n$ and $\bra m_j \ket$ are obtained by minimizing the free energy.

In the ferromagnetic state, $\bra m_j \ket$ is independent of $j$, i.e., $\bra m_j \ket=m$. 
The self-consistency equations are given by 
\bea
&&n =  \frac{1}{N}\sum_{\vk} \left[ f \left(\xi_{\vk} -Um -\frac{h}{2}\right) 
+ f\left(\xi_{\vk} +Um +\frac{h}{2}\right) \right], 
\label{n-ferro} \\
&& m =  \frac{1}{2N} \sum_{\vk} \left[ f\left(\xi_{\vk} -Um -\frac{h}{2}\right) 
- f\left(\xi_{\vk} +Um +\frac{h}{2}\right) \right] 
\label{m-ferro} \,. 
\eea
Here 
\be
\xi_{\vk} =  -2t(\cos k_x + \cos k_y) -4t'\cos k_x \cos k_y + \frac{Un}{2} -\mu \,,
\ee
and $f$, $\mu$, and $N$ are the Fermi distribution function, 
the chemical potential, and the total number of lattice sites, 
respectively, and the summation of $\vk$ is taken over the first Brillouin zone.

In the case of the antiferromagnetic state, the magnetization is described by 
$\bra m_j \ket = m_{\vQ} e^{{\rm i}{\bf Q}\cdot{\bf r}_j}$. Here $m_{\vQ}$ is 
the staggered magnetization, which is the order parameter of antiferromagnetism.  
The self-consistency equations are given by  
\bea
&&n =  \frac{2}{N} \sideset{}{'}  \sum_{\vk} \left[ f( E^+_{\vk}) + f(E^-_{\vk})\right], 
\label{n-af} \\
&&m_{\vQ} = - \frac{1}{N} \sideset{}{'}  \sum_{\vk}  
 \frac{Um_{\vQ}+\frac{h}{2}}{D_{\vk}} 
\left[ f(E^+_{\vk}) - f(E^-_{\vk})\right], 
\label{m-af}
\eea
where the summation of $\vk$ is taken over the magnetic Brillouin zone, namely  
$|k_x| + |k_y| \leq \pi$, and 
\bea
&&E^{\pm}_{\vk} = \xi^+_{\vk} \pm D_{\vk} \,, 
\label{Epm}\\
&&\xi^{\pm}_{\vk} = \frac{1}{2}(\xi_{\vk} \pm \xi_{\vk+\vQ}) \,, \\ 
&&D_{\vk} = \sqrt{(\xi^-_{\vk})^2+\left(Um_{\vQ} + \frac{h}{2}\right)^2} 
\label{Dk}\,.
\eea

A comprehensive mean-field analysis of the Hubbard model \cite{igoshev10}  
clarified the parameter region where ferromagnetic phases with $\vq=0$ 
and antiferromagnetic phases with $\vq=\vQ$ are stabilized. 
Referring to Ref.~\onlinecite{igoshev10}, we fix $U=3t$ and choose 
$t'=-0.45t$ and $n=0.2$ to describe the ferromagnetic state, and 
$t'=-0.2t$ and $n=1.1$ for the antiferromagnetic state. 
Our conclusions, however, do not depend on the  choice of parameters as long as 
the ferromagnetic (or antiferromagnetic) phase is stabilized. 
In the following, we set $t=1$ and measure all quantities with the dimensions of energy 
in units of $t$.

\section{Uniform susceptibility in the ferromagnetic state}
The longitudinal magnetic susceptibility 
is obtained in the RPA by taking a first derivative with respect to a field in Eqs.~(\ref{n-ferro}) and (\ref{m-ferro}),  
and then by taking the limit of $h\rightarrow +0$. 
One would assume that a first derivative of $\mu$ (or $n$) with respect to a field should vanish  
in the limit of $h \rightarrow +0$. This is actually correct at least in the disordered phase. 
As a result, the longitudinal susceptibility, which is defined by 
$\frac{\partial m}{\partial h}\left|_{h \to +0}\right.$,  
is obtained as 
\be
\tilde{\chi}({\bf 0}) = 
 \frac{1}{4} \frac{\chi_\uparrow+\chi_\downarrow}
 {1-\frac{U}{2}(\chi_\uparrow+\chi_\downarrow)} \,,
\label{tildechi-ferro}
\ee
where 
\be
 \chi_{\uparrow(\downarrow)} =  -\frac{1}{N}\sum_{\vk} f'(\xi_{\vk} \mp Um) 
 \label{chiup-ferro}
\ee
and $f'$ is the first derivative with respect to energy.

%%%%%{FIG.1}%%%%%%
\begin{figure}[htb]
\centering
\includegraphics[width=8.0cm]{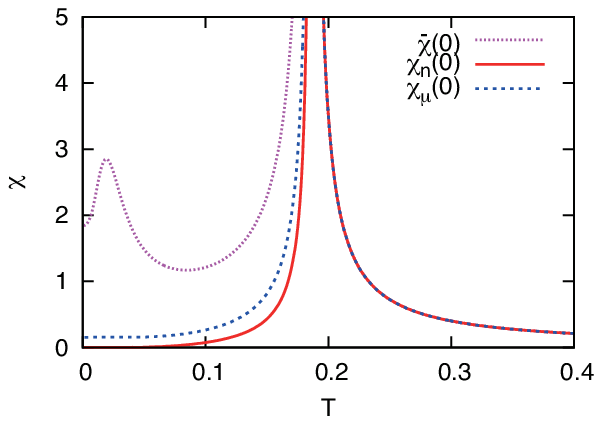}
\caption{Longitudinal magnetic susceptibility as a function of temperature 
at a fixed density; $n=0.2, t'=-0.45$, and $U=3$. 
The Curie temperature is $T_{\rm FM}=0.187$, below which the ferromagnetic moment develops. 
$\tilde{\chi}({\bf 0})$ is obtained from \eq{tildechi-ferro} and correct only in $T> T_{\rm FM}$. 
$\chi_n ({\bf 0})$ is given by \eq{chic-ferro} and correct in the whole temperature region. 
For $\chi_{\mu} ({\bf 0})$ [\eq{chigc-ferro}],    
the chemical potential is tuned at each temperature to reproduce the fixed density 
$n=0.2$. $\chi_{\mu} ({\bf 0})$ provides the correct result only in $T> T_{\rm FM}$. 
}
\label{chi-ferro}
\end{figure}
%%%%%{END  FIG.1}%%%%%

\begin{figure}[htb]
\centering
\includegraphics[width=8.0cm,clip]{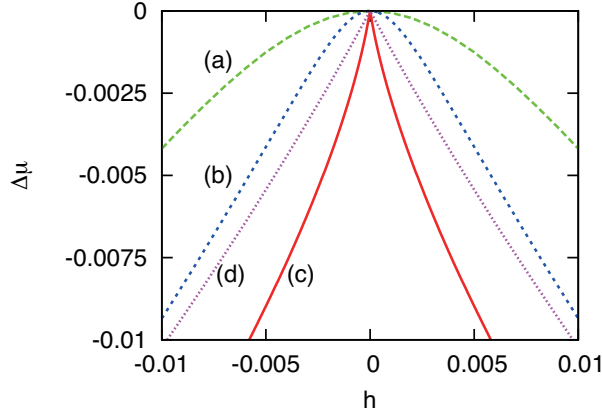}
\caption{$h$ dependence of the chemical potential for several choices of temperatures: 
(a) $T=1.1 T_{\rm FM}$, (b) $T=1.04 T_{\rm FM}$, (c) $T=0.99 T_{\rm FM}$, and 
(d) $T=0.9 T_{\rm FM}$. 
}
\label{mu-h-ferro}
\end{figure}

The temperature ($T$) dependence of  $\tilde{\chi}({\bf 0})$ is shown in \fig{chi-ferro}. 
With decreasing $T$, $\tilde{\chi}({\bf 0})$ grows and diverges at   
the Curie temperature $T_{\rm FM}$ ($= 0.187$). 
Below $T_{\rm FM}$, ferromagnetic order $m$ develops. The value of $m$ is determined by 
the self-consistency equations Eqs.~(\ref{n-ferro}) and (\ref{m-ferro}). 
As expected, $\tilde{\chi}({\bf 0})$ is suppressed below $T_{\rm FM}$. 
However, it is enhanced at lower temperature inside the ferromagnetic state. 
This dependence is obviously unphysical and originates from the wrong assumption 
that the chemical potential should remain a quadratic function with respect to a field 
inside the ferromagnetic state. 
To show this, we plot $\Delta \mu(T,h) \equiv \mu(T,h)- \mu(T,0)$ in \fig{mu-h-ferro}. 
The chemical potential $\mu$ has a quadratic dependence of $h$ in the vicinity of $h=0$ 
in the disordered phase because of the spin-rotational symmetry of the system. 
Its curvature around $h=0$ becomes larger upon approaching $T_{\rm FM}$ and 
becomes infinite just at $T_{\rm FM}$. 
Below $T_{\rm FM}$, 
a linear term emerges with a singularity at $h=0$. 
The emergence of the linear term is due to the breaking of the spin rotational symmetry, 
that is, the system has a different response when an infinitesimally small field is applied 
parallel and anti-parallel to the direction of the ferromagnetic moment. 
%The singularity at $h=0$ is due to the change of the direction of the magnetic moment 
%the direction of the magnetic moment $m$ for a positive $h$ 
%is opposite to that for negative $h$. 
%%% added %%%
%(In Fig.2, $\Delta \mu$ is an even function of $h$ even for $T<T_{FM}$. 
%This is because the magnetic moment $m$ is assumed to align with 
%the +z (-z) direction when a positive (negative) $h$ is applied. 
%%%   
%If $m$ is fixed to be positive, the response to the positive and negative 
%$h$ is completely different. In the latter case, $m$ suddenly changes 
%to the opposite direction.) 
%%%%%%%%%%
Therefore the emergent linear term in $h$ is crucially important to describe the 
response in the ordered phase and \eq{tildechi-ferro} is valid only in the disordered phase where $m=0$. 
While $\tilde{\chi}({\bf 0})$ is enhanced below $T \lesssim 0.05$ in \fig{chi-ferro} for the 
present choice of the parameters, it could diverge inside the ferromagnetic phase,  
especially when $U$  is chosen to be a larger value.

%%%%%{FIG.2}%%%%%%
%%%%%{END  FIG.2}%%%%%

\subsection{Fixed density} 
\begin{figure}[htb]
\centering
\includegraphics[width=8.0cm,clip]{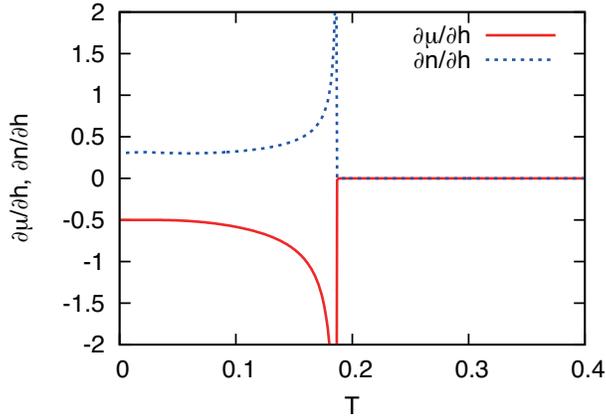}
\caption{Temperature dependence of the first derivative of $\mu$ and $n$ with respect 
to a magnetic field $h$; $T_{\rm FM}=0.187$, $n=0.2, t'=-0.45$, and $U=3$. 
}
\label{mu-n-h-ferro}
\end{figure}

We first consider the situation where the density is fixed. 
In order to get the correct RPA susceptibility inside the ordered phase, 
a first derivative of $\mu$ should be kept when differentiating 
Eqs.~(\ref{n-ferro}) and (\ref{m-ferro}) with respect to $h$. 
Solving coupled equations, we obtain 
\bea
&&\chi_{n}({\bf 0}) = \left. \frac{\partial m}{\partial h}\right|_{h \to +0}= 
\frac{\chi_\uparrow\chi_\downarrow}{\chi_\uparrow+\chi_\downarrow
-2U\chi_\uparrow\chi_\downarrow} 
\label{chic-ferro}\,,   \\
&& \left.\frac{\partial \mu}{\partial h}\right|_{h \to +0}= 
-\frac{1}{2} \frac{\chi_\uparrow - \chi_\downarrow} 
{\chi_\uparrow+\chi_\downarrow-2U\chi_\uparrow\chi_\downarrow}\,.
\label{muc-ferro}
\eea
In the disordered phase, we have $\chi_\uparrow = \chi_\downarrow$. 
Hence \eq{chic-ferro} is reduced to \eq{tildechi-ferro} and 
$\left.\frac{\partial \mu}{\partial h}\right|_{h \to +0}=0$. 
However, inside the ferromagnetic phase, it is clear that 
the functional form of \eq{chic-ferro} is very different from \eq{tildechi-ferro} 
and in addition $\left.\frac{\partial \mu}{\partial h}\right|_{h \to +0}$ becomes finite. 
We plot the temperature dependence of $\chi_{n}({\bf 0})$ in \fig{chi-ferro}. 
$\chi_{n}({\bf 0})$ is suppressed monotonically inside the ferromagnetic phase 
with decreasing temperature. This is because the system becomes less susceptible 
to an infinitesimally small field parallel to the magnetic moment when the magnetic moment 
grows with decreasing temperature. The enhancement of \eq{tildechi-ferro} inside the 
ferromagnetic phase (\fig{chi-ferro}), therefore should be 
an artifact due to the discarding of the contribution from $\left.\frac{\partial \mu}{\partial h}\right|_{h \to +0}$. 
As already implied in \fig{mu-h-ferro}, the contribution of $\left.\frac{\partial \mu}{\partial h}\right|_{h \to +0}$ 
is indeed sizable below $T_{\rm FM}$. 
The temperature dependence of $\left.\frac{\partial \mu}{\partial h}\right|_{h \to +0}$ 
is plotted in \fig{mu-n-h-ferro}. 
The quantity $\left.\frac{\partial \mu}{\partial h}\right|_{h \to +0}$  
is zero down to $T=T_{\rm FM}$. 
It diverges at $T=T_{\rm FM}$ only on the side of low temperature 
and is suppressed with decreasing $T$, 
keeping a value comparable to $\chi_{n}({\bf 0})$ at low temperature (see also \fig{chi-ferro}).

%%%%{FIG.3}%%%%%%
%%%%%{END  FIG.3%%%%%

\subsection{Fixed chemical potential} 
We now consider the situation where $\mu$ is fixed. 
In this case, we differentiate Eqs.~(\ref{n-ferro}) and (\ref{m-ferro}) 
with respect to $h$ for a fixed $\mu$. 
We then obtain 
\bea
&&\chi_{\mu}({\bf 0}) = \left. \frac{\partial m}{\partial h}\right|_{h \to +0}= 
\frac{1}{4} \frac{\chi_\uparrow + \chi_\downarrow + 2U \chi_\uparrow \chi_\downarrow } 
{1- U^2 \chi_\uparrow\chi_\downarrow}
\label{chigc-ferro}\,,   \\
&& \left.\frac{\partial n}{\partial h}\right|_{h \to +0}= 
\frac{1}{2} \frac{\chi_\uparrow - \chi_\downarrow} 
{1- U^2 \chi_\uparrow\chi_\downarrow}\,.
\label{ngc-ferro}
\eea
Equation~(\ref{chigc-ferro}) is already known in the literature \cite{izuyama63,moriya}. 
 
In the disordered phase, we have $\chi_{\uparrow} = \chi_{\downarrow}$, yielding 
$\left.\frac{\partial n}{\partial h}\right|_{h \to +0}=0$ and 
$\chi_{\mu}=\chi_n=\tilde{\chi}$. 
Consequently, the magnetic susceptibility at a fixed density is the same 
as that at a fixed chemical potential in the disordered phase. 

In the ordered phase, however, we have $\chi_{\uparrow} \neq \chi_{\downarrow}$, 
and $\left.\frac{\partial n}{\partial h}\right|_{h \to +0}$ becomes finite as 
shown in \fig{mu-n-h-ferro} and diverges at $T=T_{\rm FM}-0$. 
A comparison of Eqs.~(\ref{chic-ferro}) and (\ref{chigc-ferro}) should be made 
in the same condition, namely the same density and the same chemical potential. 
A physical quantity computed at a fixed chemical potential 
is frequently used to describe a system with a fixed density by tuning 
the chemical potential to 
reproduce the density. Following this standard procedure, we plot the temperature dependence 
of $\chi_{\mu}$ also in \fig{chi-ferro}. Although the functional forms of 
Eqs.~(\ref{chic-ferro}) and (\ref{chigc-ferro}) are different, both provide similar results in 
the ordered phase. Nevertheless, in a strict sense, $\chi_{\mu}$ does not lead to 
the correct result $\chi_n$ when the density is fixed in the system.  
Conversely, $\chi_{\mu}$ [\eq{chigc-ferro}] would provide the correct result when the 
chemical potential is fixed in the system (see Appendix). In this case, $\chi_n$ [\eq{chic-ferro}] is in turn not correct 
even if the density is tuned to reproduce the fixed chemical potential. 
The reason why the longitudinal magnetic susceptibility $\chi_{\mu}$ does not agree 
with $\chi_n$ 
in the magnetically ordered phase is that $\mu$ and $n$ are not 
symmetric in Eqs.~(\ref{n-ferro}) and (\ref{m-ferro}), and thus 
the field dependences of $\mu$ and $n$ (see \fig{mu-n-h-ferro}) 
are different from each other.

\section{Staggered susceptibility in the antiferromagnetic state} 
\begin{figure}[htb]
\centering
\includegraphics[width=8.0cm,clip]{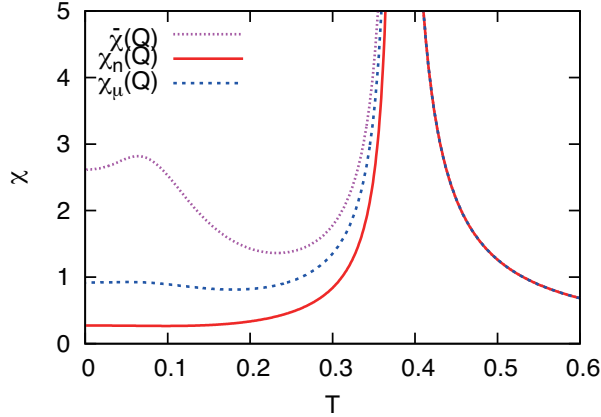}
\caption{Longitudinal magnetic susceptibility as a function of temperature 
at a fixed density; $n=1.1, t'=-0.2$, and $U=3$. 
The N\'{e}el temperature is $T_{\rm AF}=0.380$, below which the antiferromagnetic moment develops. 
$\tilde{\chi}(\vQ)$ is obtained from \eq{tildechi-af} and correct only in $T> T_{\rm AF}$. 
$\chi_n ({\vQ})$ is given by \eq{chic-af} and correct in the whole temperature region. 
$\chi_{\mu} (\vQ)$ [\eq{chigc-af}]  is computed in the condition of a fixed chemical potential; 
the chemical potential is tuned at each temperature to reproduce the correct density.  
The result $\chi_{\mu} (\vQ)$ is, however, correct only in $T> T_{\rm AF}$. }
\label{chi-af}
\end{figure}

The longitudinal staggered susceptibility is defined as 
$\left.\frac{\partial m_{\vQ}}{\partial h}\right|_{h \to +0}$ 
where $h$ is a magnitude of a staggered field introduced in \eq{zeeman} 
with $\vq=\vQ$. In the disordered phase, the spin rotational symmetry is preserved and thus 
$\mu$ and $n$ are quadratic functions of $h$ for a small $h$. In this case, we have 
$\left.\frac{\partial \mu}{\partial h}\right|_{h \to +0} =0$ and $\left.\frac{\partial n}{\partial h}\right|_{h \to +0}=0$.  
Thus we do not need to consider a first derivative of $\mu$ and $n$ with respect to $h$ in \eq{m-af}. 
The staggered susceptibility then becomes 
\be
\tilde{\chi}(\vQ) = \frac{1}{2} \frac{\chi^{(0)} (\vQ)}{1-U \chi^{(0)} (\vQ)}\,,
\label{tildechi-af}
\ee
where 
\bea
&&\chi^{(0)} (\vQ)= -\frac{1}{N} \sideset{}{'} \sum_{\vk}  \frac{(\xi^{-}_{\vk})^2}{D_{\vk}^3}
 \left( f(E^+_{\vk})-f(E^-_{\vk})\right) \nonumber \\
&& \hspace{12mm}  -\frac{1}{N} \sideset{}{'}\sum_{\vk}  \left(\frac{Um_{\vQ}}{D_{\vk}}\right)^2  
\left( f^{'}(E^+_{\vk})+f^{'}(E^-_{\vk})\right) \,, 
\label{tildechi0-af}
\eea
and $m_{\vQ}=0$ here. 
One might  apply the formula \eq{tildechi-af} to the antiferromagnetic phase, employing 
$m_{\vQ}$ and $n$ (or $\mu$)  determined by solving the self-consistency equations 
Eqs.~(\ref{n-af}) and (\ref{m-af}). 
The resulting $\tilde{\chi}(\vQ)$ [\eq{tildechi-af}] is shown in \fig{chi-af} as a function of temperature. 
With decreasing $T$, $\tilde{\chi}(\vQ)$ grows and diverges at the N{\'e}el temperature 
$T_{\rm AF}=0.380$. Just below $T_{\rm AF}$, $\tilde{\chi}(\vQ)$ is suppressed 
as expected. However, it grows below $T \lesssim 0.2$ 
inside the antiferromagnetic phase. 
This apparently unphysical 
result originates from the wrong assumption that 
$\mu$ and $n$ would still be quadratic in $h$ in the magnetic phase. 
The enhancement of  $\tilde{\chi}(\vQ)$ could appear as its divergence 
at a certain temperature below $T_{\rm AF}$ when a larger value of $U$ is taken.  

%%%%%{FIG.4}%%%%%%
%%%%%{END  FIG.4}%%%%%

\begin{figure}[htb]
\centering
\includegraphics[width=8.0cm,clip]{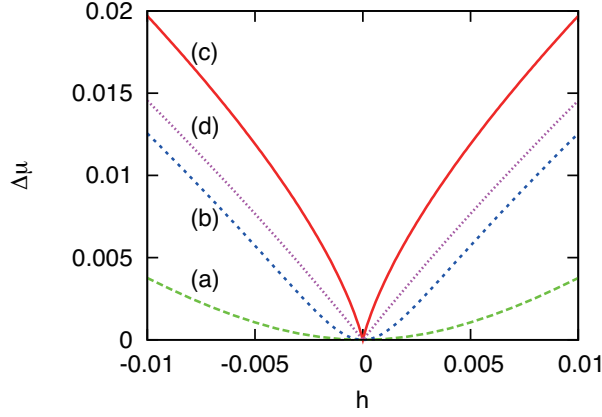}
\caption{$h$ dependence of the chemical potential for several choices of temperatures: 
(a) $T=1.1 T_{\rm AF}$, (b) $T=1.03 T_{\rm AF}$, (c) $T=0.99 T_{\rm AF}$, and (d) $T=0.9 T_{\rm AF}$; 
$T_{\rm AF}=0.380$, $n=1.1, t'=-0.2$ and $U=3$. 
}
\label{mu-h-af}
\end{figure}

Figure~\ref{mu-h-af} shows $\Delta\mu=\mu(T,h) - \mu(T,0)$ as a function of $h$ 
for several choices of $T$. For $T> T_{\rm AF}$ we see 
$\left.\frac{\partial \mu}{\partial h}\right|_{h \to +0}=0$. 
However, below $T_{\rm AF}$, $\Delta\mu$ becomes singular at $h=0$ and 
acquires a linear dependence of $|h|$ around $h=0$. 
This effect is crucially important to obtain the correct RPA expression 
of the longitudinal magnetic susceptibility inside the magnetic phase. 
Because the correct expression depends on whether the density is 
fixed or the chemical potential is fixed, we present it below separately.

%%%%{FIG.5}%%%%%%
%%%%%{END  FIG.5}%%%%%
%

\subsection{Fixed density} 
\begin{figure}[htb]
\centering
\includegraphics[width=8.0cm,clip]{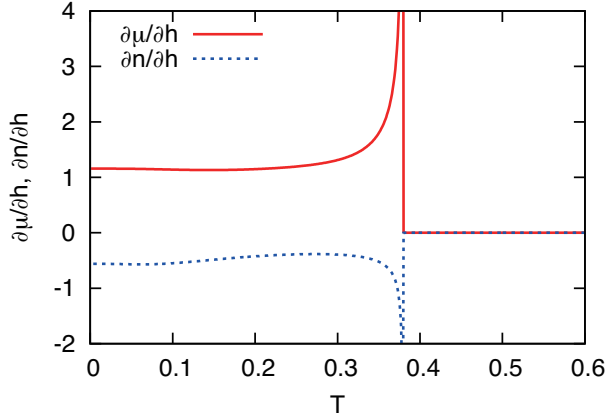}
\caption{Temperature dependence of the first derivative of $\mu$ and $n$ with respect 
to a magnetic field $h$; $T_{\rm AF}=0.380$, $n=1.1, t'=-0.2$, and $U=3$. 
}
\label{mu-n-h-af}
\end{figure}

For a fixed density $n$, we differentiate both Eqs.~(\ref{n-af}) and (\ref{m-af}) with respect to $h$ 
and take the limit of $h\rightarrow +0$. Coupled equations of 
$\frac{\partial m}{\partial h}$ and $\frac{\partial \mu}{\partial h}$ 
are easily solved, yielding 
\bea
&&\chi_{n}(\vQ) = \left. \frac{\partial m}{\partial h}\right|_{h \to +0}= 
\frac{1}{2} 
\frac{\chi_{n}^{(0)}(\vQ)}{1-U \chi_{n}^{(0)}(\vQ)} 
\label{chic-af}\,,   \\
&& \left.\frac{\partial \mu}{\partial h}\right|_{h \to +0}= 
-\frac{1}{2} \frac{a_{12} / a_{11}} {1-U \chi_{n}^{(0)}(\vQ)} \,,
\label{muc-af}
\eea
where 
\be
\chi_{n}^{(0)}(\vQ) = \frac{a_{11} a_{22} - a_{12} a_{21}}{ a_{11}} \,,
\label{chic0-af}
\ee
and 
\bea
&& a_{11} = -\frac{2}{N} \sideset{}{'}\sum_{\vk} \left( 
f^{'}(E^+_{\vk})+f^{'}(E^-_{\vk}) \right) \,,\\
&& a_{12}= \frac{2}{N}  \sideset{}{'}\sum_{\vk} 
\frac{Um_{\vQ}}{D_{\vk}} \left( 
f^{'}(E^+_{\vk})-f^{'}(E^-_{\vk}) \right) \,,\\
&& a_{21}= \frac{1}{2} a_{12} \,, \\
&& a_{22}= \chi^{(0)}(\vQ) \,.
\label{a22} 
\eea
Here $h$ should be put zero in $E^{\pm}_{\vk}$ and $D_{\vk}$ [see Eqs.~(\ref{Epm}) and  (\ref{Dk})]. 
The functional form of \eq{chic-af} is the same as \eq{tildechi-af}, but 
$\chi^{(0)}_n (\vQ)$ becomes identical to $\chi^{(0)}(\vQ)$ only for $m_{\vQ}=0$. 
$\chi_n(\vQ)$ is plotted in \fig{chi-af} as a function of temperature. 
It is the same as \eq{tildechi-af} above $T_{\rm AF}$. 
Below $T_{\rm AF}$, $\chi_n(\vQ)$ is suppressed monotonically with decreasing temperature 
as it should be. 
In \fig{mu-n-h-af} we plot $\left.\frac{\partial \mu}{\partial h}\right|_{h \to +0}$. 
It vanishes in the disordered phase, but becomes sizable in the magnetically ordered phase 
with divergence at $T=T_{\rm AF}$ on the side of low temperature.

%%%%{FIG.6}%%%%%%
%%%%%{END  FIG.6}%%%%%
%

\subsection{Fixed chemical potential} 
We next fix the chemical potential and differentiate Eqs.~(\ref{n-af}) and (\ref{m-af}) 
with respect to $h$. Taking the limit of $h\rightarrow +0$, we obtain 
\bea
&&\chi_{\mu}(\vQ) = \left. \frac{\partial m}{\partial h}\right|_{h \to +0}= 
\frac{1}{2} 
\frac{\chi_{\mu}^{(0)}(\vQ)}{1-U \chi_{\mu}^{(0)}(\vQ)} 
\label{chigc-af}\,,   \\
&& \left.\frac{\partial n}{\partial h}\right|_{h \to +0}= 
\frac{1}{2\left( 1+\frac{U}{2} a_{11}\right)} \frac{a_{12}} {1-U \chi_{\mu}^{(0)}(\vQ)} \,,
\label{ngc-af}
\eea
where 
\be
\chi^{(0)}_{\mu}(\vQ) = a_{22} -\frac{U}{2} \frac{a_{12} a_{21}}{1+\frac{U}{2} a_{11}} \,. 
\label{chigc0-af}
\ee
The functional form of \eq{chigc-af} is the same as \eq{chic-af} obtained at a fixed density. 
However, the expression of $\chi^{(0)}_{\mu}$ is very different from 
$\chi^{(0)}_{n}$ [\eq{chic0-af}]. They become the same only in the disordered phase, 
where $m_{\vQ}=0$ and thus $a_{12}=a_{21}=0$. 
Temperature dependence of $\chi_{\mu}(\vQ)$ is shown in \fig{chi-af}.  
Since the density is fixed in \fig{chi-af}, the chemical potential 
is tuned to reproduce the correct density at each temperature,  
as is usually done. Below $T_{\rm AF}$, $\chi_{\mu}(\vQ)$ is suppressed, but 
does not reproduce the correct result of $\chi_{n}(\vQ)$. 
This wrong result originates from the naive assumption that 
the susceptibility obtained at a fixed chemical potential could be used 
for the system with a fixed density after tuning the chemical potential to reproduce 
the correct density. 
However, as we have obtained explicitly, 
the susceptibility at a fixed chemical potential [Eqs.~(\ref{chigc-af}) and (\ref{chigc0-af})] 
is different from that at a fixed density [Eqs.~(\ref{chic-af}) and (\ref{chic0-af})-(\ref{a22})] 
in the magnetically ordered phase. Furthermore, as shown in \fig{mu-n-h-af}, 
temperature dependence of 
$\left.\frac{\partial n}{\partial h}\right|_{h \to +0}$ is very different from 
$\left.\frac{\partial \mu}{\partial h}\right|_{h \to +0}$. 
Therefore the choice of $\chi_n$ and $\chi_{\mu}$ should be made carefully 
to describe the system appropriately.  
Reversely, if we wish to describe the system with a fixed chemical potential, 
the susceptibility $\chi_{\mu}(\vQ)$ is the correct one and $\chi_{n}(\vQ)$ [\eq{chic-af}] 
does not reproduce the correct result even if the density is tuned to reproduce the correct 
chemical potential at each temperature (see Appendix).

\section{Diagrammatic approach} 
It is  natural to ask what kind of result is obtained when a diagrammatic approach 
is employed. The longitudinal magnetic susceptibility is defined by 
\be
\chi^{zz}(\vq, {\rm i}\omega_m) = \frac{1}{N} \int_{0}^{1/T} {\rm d}\tau 
{\rm e}^{{\rm i} \omega_m \tau} \bra T_{\tau}  S^{z}(\vq,\tau) S^{z}(-\vq,0) \ket \,,
\ee
where $\omega_m=2m\pi T$ is the bosonic Matsubara frequency with $m$ being 
integer, $S^{z}(\vq,\tau)={\rm e}^{\tau \mathcal{H}} S^{z} (\vq) {\rm e}^{-\tau \mathcal{H}} $, 
and $S^{z}(\vq)=\frac{1}{2} \sum_{\vk \sigma} \sigma c_{\vk \sigma}^{\dagger} c_{\vk+\vq \sigma}$. 

\begin{figure}[htb]
\centering
\includegraphics[width=12.0cm,clip]{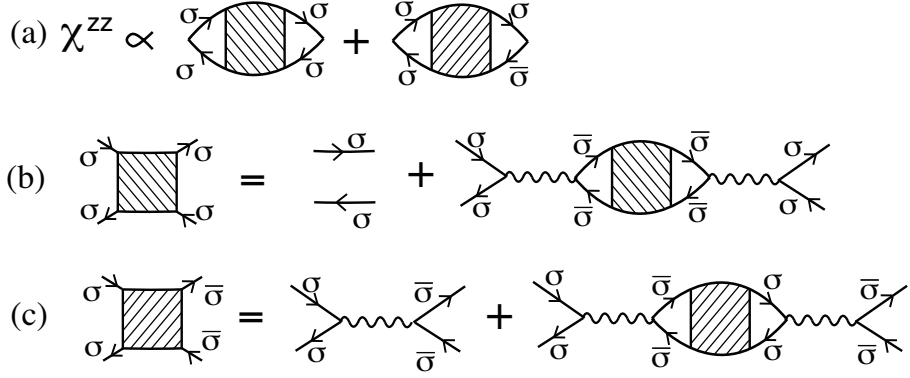}
\caption{Diagrams of the longitudinal magnetic susceptibility for the Hubbard interaction. 
$\bar{\sigma}$ denotes the spin direction opposite to $\sigma$. 
}
\label{diagram}
\end{figure}

In the disordered phase, $\chi^{zz}$ is given by the diagrams shown in \fig{diagram} in the RPA. 
Hence we obtain 
\be
\chi^{zz} =  \frac{1}{4} \left( 
\frac{\chi_{\uparrow} + \chi_{\downarrow}} {1-U^2 \chi_{\uparrow} \chi_{\downarrow}} 
+ \frac{2U \chi_{\uparrow} \chi_{\downarrow}} {1-U^2 \chi_{\uparrow} \chi_{\downarrow}} 
\right) \,.
\label{chizz}
\ee
In the static case, we set ${\rm i} \omega_m =0$ and take $\vq={\bf 0}$ and $\vQ$ 
for the uniform and staggered susceptibility, respectively. 
We then obtain $\chi_{\uparrow}=\chi_{\downarrow}=\chi_0$, which is 
the same as \eq{chiup-ferro} 
for the uniform susceptibility and \eq{tildechi0-af} 
for the staggered susceptibility. 
Hence \eq{chizz} is reduced to 
\be
\chi^{zz} = \frac{1}{2} \frac{\chi_0}{1-U \chi_0} \,,
\ee
and we reproduce the correct results Eqs.~(\ref{tildechi-ferro}) and (\ref{tildechi-af}) 
in the disordered phase. 

%%%%{FIG.7}%%%%%%
%%%%%{END  FIG.7}%%%%%

In the ordered phase, we may compute $\chi_{\uparrow (\downarrow)}$ by 
using the quasiparticle propagator. In the ferromagnetic phase, 
\eq{chizz} then becomes the same as $\chi_{\mu}({\bf 0})$ [see \eq{chigc-ferro} 
and Refs.~\onlinecite{izuyama63} and \onlinecite{moriya}], 
but not $\chi_{n}(\bf 0)$ [\eq{chic-ferro}]. 

The situation is delicate in the antiferromagnetic phase. 
Although the translational symmetry is broken by the magnetic order, 
the umklapp components of the susceptibility 
such as $\bra T_{\tau}  S^{z}(\vq,\tau) S^{z}(-\vq-\vQ,0) \ket$ and 
$\bra T_{\tau}  S^{z}(\vq+\vQ,\tau) S^{z}(-\vq,0) \ket$, do not contribute to the longitudinal 
susceptibility \cite{schrieffer89}. Hence one might think that the RPA susceptibility 
would be obtained simply by replacing the electron Green's function with 
the Green's function of the two-component field 
$\psi_{\vk \sigma}^{\dagger} = (c_{\vk \sigma}^{\dagger}, c_{\vk +\vQ \sigma}^{\dagger})$; 
the summation of $\vk$ is then restricted to the magnetic Brillouin zone. 
This kind of calculation is frequently seen in 
the literature \cite{sokoloff69,schrieffer89, kampf94,knolle10,rowe12}. 
In this case, however, we obtain $\chi_{\uparrow} (\vQ) = \chi_{\downarrow} (\vQ) = 
\chi^{(0)}(\vQ)$ [see \eq{tildechi0-af}],  
which is the same as $\tilde{\chi}(\vQ)$ and does not reproduce the correct result inside the 
antiferromagnetic phase as we have seen in \fig{chi-af}. 
The correct procedure \cite{chubukov92,misc-chubukov92,hjlee12} is to take 
into account the umklapp components such as 
$\bra T_{\tau}  S^{z}(\vq,\tau) \rho(-\vq-\vQ,0) \ket$ as well as 
the density fluctuations with $\vq+\vQ$, namely 
$\bra T_{\tau}  \rho(\vq+\vQ,\tau) \rho(-\vq-\vQ,0) \ket$. 
The density operator may be given by 
$\rho(\vq)=\frac{1}{2} \sum_{\vk}\sum_{\sigma} c_{\vk \sigma}^{\dagger}c_{\vk+\vQ \sigma}
=\frac{1}{2}  \sum_{\vk}^{'} \sum_{\sigma} \psi_{\vk \sigma}^{\dagger} \psi_{\vk+\vQ \sigma}$, 
where the factor of $1/2$ is added to make the formalism simpler. 
The resulting RPA expression becomes 
\be
\hat{\chi} = \left( 1- \hat{\chi_0} \hat{U} \right)^{-1} \hat{\chi_{0}} \,,
\ee
and 
\bea
&&\hat{\chi}= \left(
\begin{array}{cc}
\chi^{zz}(\vq,{\rm i}\omega_m) & \chi^{z\rho}(\vq, \vq+\vQ, {\rm i}\omega_m)\\
\chi^{\rho z}(\vq+\vQ, \vq, {\rm i}\omega_m) & \chi^{\rho \rho}(\vq+\vQ, {\rm i}\omega_m)
\end{array}
\right)\,, \\
&&
\hat{U}= \left(
\begin{array}{cc}
2U & 0\\
0 & -2U
\end{array}
\right)\,.
\eea
Here 
\bea
&&\chi^{z\rho}(\vq, \vq+\vQ, {\rm i}\omega_m) = \frac{1}{N} \int_{0}^{1/T} {\rm d}\tau 
{\rm e}^{{\rm i} \omega_m \tau} \bra T_{\tau}  S^{z}(\vq,\tau) \rho(-\vq-\vQ,0) \ket \,,\\
&&\chi^{\rho z}(\vq+\vQ, \vq, {\rm i}\omega_m) = \frac{1}{N} \int_{0}^{1/T} {\rm d}\tau 
{\rm e}^{{\rm i} \omega_m \tau} \bra T_{\tau} \rho(\vq+\vQ,\tau) S^{z}(-\vq,0) \ket \,,\\
&&\chi^{\rho \rho}(\vq+\vQ, {\rm i}\omega_m) = \frac{1}{N} \int_{0}^{1/T} {\rm d}\tau 
{\rm e}^{{\rm i} \omega_m \tau} \bra T_{\tau}  \rho(\vq+\vQ,\tau) \rho(-\vq-\vQ,0) \ket \,, 
\eea
and $\hat{\chi_0}$ denotes a bare susceptibility matrix where 
each element is given by a simple bubble diagram.  
Setting  ${\rm i}\omega_m =0$ and $\vq=\vQ$,  we obtain $\chi^{zz}(\vQ,0)$, which reproduces 
\eq{chigc-af}. That is, the effect of $\frac{\partial n}{\partial h}$ is 
taken into account diagrammatically by considering the contribution from 
the density-density interaction such as $\chi^{z \rho}$,  $\chi^{\rho z}$, and  $\chi^{\rho \rho}$. 

The diagrammatic method is formulated in the grand canonical ensemble.  
Hence it is natural that we can successfully reproduce both results $\chi_{\mu}({\bf 0})$ [\eq{chigc-ferro}] 
in the ferromagnetic phase and $\chi_{\mu}(\vQ)$ [\eq{chigc-af}] in the antiferromagnetic phase. 
A remaining problem is how to reproduce $\chi_{n}({\bf 0})$ [\eq{chic-ferro}]  
and $\chi_{n}({\vQ})$ [\eq{chic-af}] obtained at a fixed density in terms of the diagrammatic method. 
As we have shown explicitly in Sec. III and IV, 
the longitudinal magnetic susceptibility at a fixed density is 
different from that at a fixed chemical potential in the magnetically ordered phase. 
Given that the density is usually fixed in the actual material, it is an important problem 
to find a general recipe to compute the magnetic susceptibility in the ordered phase 
at a fixed density.

\section{Concluding remarks}
We have studied the longitudinal magnetic susceptibility by employing 
the two-dimensional Hubbard model. 
In the magnetically ordered phase, the spin rotational symmetry is broken 
and thus $\mu$ and $n$ acquire a linear term in a magnetic field when 
the field is applied parallel to the direction of the magnetic moment. 
Because of this effect, a careful analysis is required: the longitudinal magnetic 
susceptibility becomes different when computed at a fixed density and  a fixed 
chemical potential. We have provided the correct 
expressions Eqs.~(\ref{chic-ferro}) and (\ref{chic-af}) at a fixed density 
and Eqs.~(\ref{chigc-ferro}) and (\ref{chigc-af}) at a fixed chemical potential 
in both ferromagnetic and antiferromagnetic states. 
It should be noted that the susceptibility obtained at a fixed chemical potential (density) 
cannot be applied to the system with a fixed density (chemical potential) 
even though the chemical potential (density) is tuned to reproduce the correct 
density (chemical potential).

While we have exemplified our issue by employing the two-dimensional Hubbard model 
in the RPA, we believe that our conclusions do not depend on the 
choice of models, dimensions, 
lattices, and approximations even beyond the RPA. This consideration is based on  thermodynamics. 
As in the case of the relation between specific heat at constant volume and that at constant 
pressure, we can derive the following relation from the thermodynamic principle: 
\be
\chi_{n} = \chi_{\mu} + \left.\frac{\partial n}{\partial h} \right|_{\mu} 
\left.\frac{\partial \mu}{\partial h} \right|_{n}   \,. 
\label{chic-chigc}
\ee
In addition, one can easily show 
that the second term in \eq{chic-chigc} becomes negative semidefinite 
and thus $\chi_n \leq \chi_{\mu}$. 
This is because 
$\left.\frac{\partial n}{\partial h} \right|_{\mu}  \left.\frac{\partial \mu}{\partial h} \right|_{n} 
= - ( \left.\frac{\partial n}{\partial h} \right|_{\mu} )^{2} 
\left.\frac{\partial \mu}{\partial n} \right|_{h}$  and 
the stability of the thermodynamic potentials indicates that 
$\left.\frac{\partial \mu}{\partial n} \right|_{h}$ should be positive semidefinite. 
Our obtained results in 
Figs.~\ref{chi-ferro}, \ref{mu-n-h-ferro}, \ref{chi-af}, and \ref{mu-n-h-af} 
indeed satisfy \eq{chic-chigc} 
numerically and we can also check analytically that 
Eqs.~(\ref{chic-ferro}), (\ref{muc-ferro}), (\ref{chigc-ferro}) and (\ref{ngc-ferro}), 
and Eqs.~(\ref{chic-af}), (\ref{muc-af}), (\ref{chigc-af}) and (\ref{ngc-af}) 
fulfill \eq{chic-chigc} in the ferromagnetic and the antiferromagnetic case, respectively. 
Moreover, Figs.~\ref{mu-n-h-ferro} and \ref{mu-n-h-af} indeed show that 
$\left.\frac{\partial n}{\partial h} \right|_{\mu}$ and $\left.\frac{\partial \mu}{\partial h} \right|_{n}$ 
have opposite signs. 
The thermodynamic relation \eq{chic-chigc} is, however, not well recognized in 
the literature. In fact, the contributions from $\frac{\partial n}{\partial h}$ and 
$\frac{\partial \mu}{\partial h}$ are frequently missed and an inappropriate 
formula such as Eqs.~(\ref{tildechi-ferro}) and (\ref{tildechi-af}) is employed 
to compute the longitudinal magnetic susceptibility in the magnetically ordered phase 
\cite{sokoloff69,schrieffer89,kampf94,knolle10,rowe12}.

As we have discussed in 
%Sec.~III and IV, 
Sec.~III (\fig{chi-ferro}) and IV (\fig{chi-af}), 
the enhancement of $\tilde{\chi}(\bf 0)$ [\eq{tildechi-ferro}] and $\tilde{\chi}(\vQ)$ [\eq{tildechi-af}] 
at low temperature inside the magnetic state is not a signal of some instability, but 
just an artifact due to the employment of the wrong susceptibility. 
Mathematically this enhancement comes from a slight enhancement of  $\chi_{\uparrow} + 
\chi_{\downarrow}$ [\eq{chiup-ferro}] and  $\chi^{(0)}(\vQ)$ [\eq{tildechi0-af}] 
due to the development of magnetic order. 
This subtle change is removed by including the effect of 
$\frac{\partial\mu}{\partial h}|_{h \to 0}$ 
in Eqs.~(\ref{chic-ferro}) and (\ref{chic-af}) or 
$\frac{\partial n}{\partial h}|_{h \to 0}$ in Eqs.~(\ref{chigc-ferro}) and (\ref{chigc-af})  
in the mean-field theory of the Hubbard model. 
However, it should be noted that in a more general situation, 
an enhancement of the susceptibility inside the magnetic phase could occur even if 
the effect of $\frac{\partial\mu}{\partial h}|_{h \to 0}$ (or $\frac{\partial n}{\partial h}|_{h \to 0}$) 
is correctly taken into account. For example, with decreasing temperature inside the magnetic phase, 
there could occur a tendency of a reentrant transition to a normal phase or a continuous transition 
to a different ordered phase.

Our obtained results are relevant to metallic systems whenever 
$\mu$ and $n$ acquire a linear dependence on a magnetic field.  
The presence of the linear term in the field is easily recognized by symmetry. 
The ferromagnetic (antiferromagnetic) state is not symmetric with respect to 
the change of the field direction, namely $h \leftrightarrow -h$ when $h$ is applied parallel 
to the direction of the uniform (staggered) magnetism. 
Therefore, we expect $\frac{\partial \mu}{\partial h} \ne 0$ and $\frac{\partial n}{\partial h} \ne 0$.  
On the other hand,  for the uniform longitudinal 
magnetic susceptibility inside the antiferromagnetic state, 
we have $\frac{\partial \mu}{\partial h}=0$ and $\frac{\partial n}{\partial h}=0$, 
because the system is symmetric with respect to the change of the direction of a uniform 
field inside the antiferromagnetic phase. 
Another example is the case of the transverse field $h_{\perp}$: 
the system is symmetric with respect to the change of the field direction 
in both the ferromagnetic and antiferromagnetic state, leading to 
$\frac{\partial \mu}{\partial h_{\perp}}=0$ and $\frac{\partial n}{\partial h_{\perp}}=0$.  
Hence the transverse magnetic susceptibility is computed without considering 
possible contributions from $\frac{\partial \mu}{\partial h_{\perp}}$ and $\frac{\partial n}{\partial h_{\perp}}$ 
as seen in the literature \cite{izuyama63,ueda78,schrieffer89, chubukov92,kampf94,knolle10,hjlee12,rowe12}. 

For an insulating state, special care may not be needed, because 
$\frac{\partial n}{\partial h}$ should vanish in \eq{chic-chigc} due to the presence of a charge gap and 
we obtain $\chi_{n}=\chi_{\mu}$. In fact, 
in an antiferromagnetic insulating state, we would have 
$E^{+}_{\vk} > 0$ and $E^{-}_{\vk} < 0$ independent of $\vk$. 
We can then easily obtain $n=1$ [\eq{n-af}]  and 
$\frac{\partial n}{\partial h}=0$ [\eq{ngc-af}] at $T=0$.  

As a direct test of the present theory, we propose a susceptibility measurement 
in two different conditions, i.e., for a fixed density and a fixed chemical potential. 
Whereas the former condition is easily controlled in experiments, 
the latter condition may require the state-of-the-art technique in which 
a magnetic metal touches a charge reservoir, for example, exploiting a field-effect 
transistor. As seen in Figs.~\ref{chi-ferro}, \ref{chi-af} and \ref{chi-mu-const}, 
we predict a sizable difference between $\chi_{n}$ and $\chi_{\mu}$ in a 
magnetically ordered phase.

\begin{acknowledgments}
The authors thank Y. Hasegawa for fruitful discussions at an early stage of the present work 
and P. Jakubczyk for critical reading of the manuscript. 
They are indebted also to Y. Kuramoto, M. Hayashi, and K. Miyake for encouraging discussions 
to pursue the present issue. 
H.Y. acknowledges support by JSPS KAKENHI Grant No. 15K05189. 
\end{acknowledgments}

\appendix
\section{System with a fixed chemical potential} 
\begin{figure}[htb]
\centering
\includegraphics[width=14.0cm,clip]{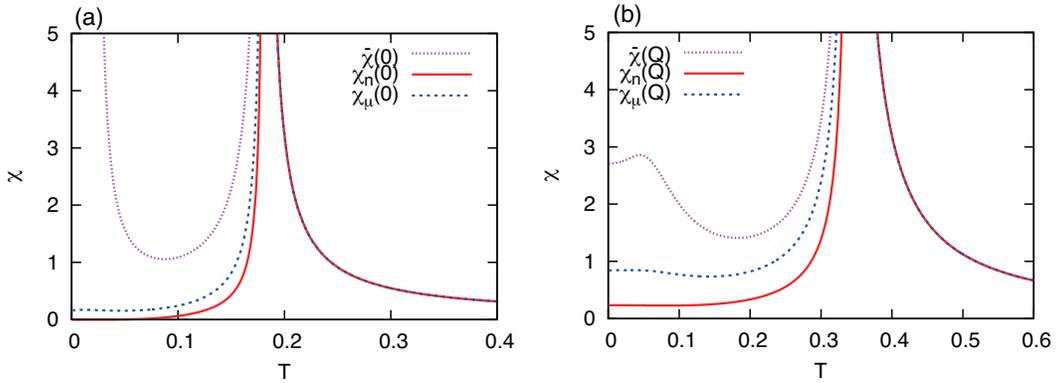}
\caption{Longitudinal magnetic susceptibility as a function of temperature for a fixed 
chemical potential in the ferromagnetic (a) and antiferromagnetic (b) phase. 
The chemical potential is chosen as $\mu=-1.83$ in (a) which reproduces $n=0.2$ at $T=0.15$,  
and as $\mu=1.69$ in (b) where $n=1.1$ at $T=0.2$. 
The other parameters are $U=3$, $t'=-0.45$ in (a) and $t'=-0.2$ in (b). 
For $\chi_n$, the density is tuned at each temperature to reproduce the fixed chemical potential. 
The result of $\chi_{\mu}$ is correct in the whole temperature region.} 
\label{chi-mu-const}
\end{figure}

The temperature dependence of the magnetic susceptibility is obtained 
for a fixed density in Figs.~\ref{chi-ferro} and \ref{chi-af}. 
Hence $\chi_{n}$ provides the correct result. 
While the density does not change as a function of temperature 
in actual materials, one can still consider a situation in which a system 
comes into contact with a charge reservoir. 
For example, a system is described as having several bands crossing the Fermi energy,  
and there is essentially only one active band with a large density of states. 
In that case, we may focus on such a band and invoke a condition of a fixed chemical 
potential. The temperature dependence of the magnetic susceptibility 
for a fixed chemical potential is shown in \fig{chi-mu-const}(a) and (b) 
in the ferromagnetic and antiferromagnetic case, respectively. 
These results are very similar to the results for a fixed density shown in 
Figs.~\ref{chi-ferro} and \ref{chi-af}. However, the correct result here is 
$\chi_{\mu}$, not $\chi_n$. 

%%%%{FIG.8}%%%%%%
%%%%%{END  FIG.9}%%%%%

\bibliography{main} 

\begin{thebibliography}{15}
\bibitem[{mis()}]{misc-chubukov92}
\bibinfo{note}{In the numerator of Eq.~(52) in Ref.~\onlinecite{chubukov92},
  $+U [\chi_{\vQ}^{z \rho}(\vq,\omega)]^{2}$ should be $-U [\chi_{\vQ}^{z
  \rho}(\vq,\omega)]^{2}$.}
\end{thebibliography}
%----------------------------------------------------------------------------------------------------------
\end{document}